\documentstyle[12pt]{article}
\newcommand{\beq}{\begin{equation}}
\newcommand{\eeq}{\end{equation}}
\newcommand{\bea}{\vspace{0.25cm}\begin{eqnarray}}
\newcommand{\eea}{\end{eqnarray}}

\newcommand{\gb}{\mbox{{\boldmath
$\gamma$}}}
\newcommand{\ab}{\mbox{{\boldmath
$\alpha$}}}
\newcommand{\r}{\mbox{{\boldmath
$\rho$}}}

\newcommand{\pb}{\mbox{{\bf
p}}}

\newcommand{\kb}{\mbox{{\bf
k}}}
\def\lsim{\mathrel{\rlap{\lower4pt\hbox{\hskip1pt$\sim$}}
    \raise1pt\hbox{$<$}}}         
\def\gsim{\mathrel{\rlap{\lower4pt\hbox{\hskip1pt$\sim$}}
    \raise1pt\hbox{$>$}}}         
\begin{document}
\thispagestyle{empty}

\begin{center}

  {\large\bf
COHERENT FINAL STATE INTERACTION IN JET PRODUCTION
IN NUCLEUS-NUCLEUS COLLISIONS
\\
\vspace{1.5cm}
  }
\medskip
  {\large
  B.G. Zakharov
  \bigskip
  \\
  }
{ L.D. Landau Institute for Theoretical Physics,\\
        GSP-1, 117940, Kosygina Str. 2, 117334 Moscow, Russia
\vspace{2.7cm}\\}

  {\bf
  Abstract}
\end{center}
{
\baselineskip=9pt
We study the coherent final state interaction of an energetic parton produced
in $AA$ collisions caused by the change in the cutoff scale and 
running coupling constant from the vacuum to QGP.
We demonstrate that the contribution of this new mechanism to the energy 
loss may be of the order of magnitude of the induced gluon radiation.
However, an accurate evaluation of this medium effect is a difficult task
because there is a strong cancellation between the cutoff and running coupling
constant effects. The uncertainties in the contribution of the coherent
final state interaction restrict strongly the accuracy of jet
tomographic analyses of the matter density produced in $AA$
reactions.
}
\pagebreak
\newpage

{\bf 1. Introduction.}
In recent years there has been much work done on the energy loss
of fast partons in hot QCD medium due to gluon radiation induced 
by multiple scattering (for a review, see \cite{BSZ}). This
is of great importance for understanding final state interaction 
in hard reactions in high energy nucleus-nucleus collisions 
which are under active investigation at RHIC, and will be 
studied in future experiments at LHC.

The theoretical calculations show that the energy loss 
in quark-gluon plasma (QGP) exceeds considerably the one in hadronic
medium \cite{BDMS,Z1,GLV1}. Since gluon radiation softens the parton 
fragmentation functions of energetic partons produced in hard reactions 
in the initial stage of $AA$ collisions it should lead to significant 
suppression of the high $p_{T}$ hadronic spectra in $AA$ collisions 
with respect to $pp$ collisions  (so-called jet quenching) if a hot QGP is 
formed \cite{Wang,LS}. 
Such a suppression was indeed recently discovered by the PHENIX experiment 
\cite{PHENIX} at RHIC for $\pi^{0}$ spectra at $p_{T}\lsim 4$ GeV in 
central $Au+Au$ collisions at $\sqrt{s}=130$ GeV.
Because the energy loss is sensitive to the density of hot medium
it looks quite natural to use experimental data on high $p_{T}$
spectra for jet tomographic analysis of the matter density produced in $AA$
reactions \cite{GLV2,SW,WW}.

To understand the range of uncertainty in the jet tomographic analyses
it is important to study other possible final state interaction effects
in jet production.
One mechanism of potential interest is 
the in-medium modification of the parton cascade without  gluon exchanges
between the fast partons and thermal partons (this mechanism we call
the coherent final state interaction (CFSI)). The reason is evident: 
jet splitting 
in vacuum is the major mechanism of the energy loss of energetic partons, 
and if the medium affects the parton cascading one can expect a significant 
modification of the fragmentation functions. For example, such a modification 
should inevitably arise as a mass effect due to different infrared cutoff 
scales in the vacuum and QGP.
Another obvious source of the CFSI is 
the in-medium modification of the running coupling constant 
$\alpha_s(k)$.\footnote{Note that these CFSI 
effects differ from the coherent double gluon exchanges which are usually
included in the induced gluon radiation to insure the unitarity 
\cite{Z1}.} 
Although at large virtualities the running coupling constants in 
the vacuum and QGP are close to each other this should not be the case at 
low $k$ where different background environments in which 
the gluon bremsstrahlung occurs can lead to a difference in the 
running coupling
constants in these two cases.
Note that both these medium effects 
should lead to transition radiation very similar to that of
photon radiation in QED.
The purpose of the present work is to address the CFSI for RHIC conditions
within a simple model for gluon radiation which will be
discussed in detail below.

{\bf 2. Cutoff scales and running coupling constants.} 
Let us first discuss the magnitudes of the cutoff scales for parton 
splitting in the vacuum and QGP.
In the QCD vacuum the natural cutoff is the inverse
gluon correlation radius $R_{c}^{-1}\sim 0.8-1$ GeV \cite{Shuryak1,Shuryak2}. 
Namely at such
a virtuality scale the perturbative cascade stops in the Monte Carlo programs
like JETSET, and the string fragmentation takes over. Note also that 
introduction of the effective gluon mass $m_{g,v}\sim R_{c}^{-1}$ 
(hereafter we use 
the index $v$ for vacuum quantities, and for the plasma quantities below we 
use the index $p$) allows one to describe
the HERA data on the low-$x$ proton structure function \cite{NZ}.
The analysis of inclusive radiative decays of the $J/\Psi$ and $\Upsilon$
\cite{Field} also gives $m_{g,v}\sim 0.7-1.2$ GeV.

In the QGP phase the nonperturbative  fluctuations are suppressed, and
the natural cutoff for radiation of transverse gluons, which propagate
through QGP as quasiparticles, is the thermal gluon mass.
At high temperature it reads
$
m_{g,p}^{2}=\frac{g^{2}T^{2}}{2}\left(\frac{N_c}{3}+
\frac{N_f}{6}\right)\,.
$
An analysis of the results of the lattice calculations show 
that in the temperature range $T\sim (1-3)T_c$ ($T_c\approx 170$ MeV 
is the temperature of the confinement phase transition), 
which is of relevance to $AA$ collisions at RHIC, 
the nonberturbative effects are still important \cite{LH}. 
Using a quasiparticle picture with massive gluons and 
quarks in the above temperature window the authors of Ref. \cite{LH}
obtained $m_{g,p}\approx 0.4$ GeV, and $m_{q,p}\approx 0.3$ GeV.
Thus we see that there is a considerable difference in the cutoffs 
in the vacuum and QGP. 

Let us recall now the situation with the running coupling constant
at low $k$. There are some indications that in 
the vacuum nonperturbative effects stop the growth of the running 
coupling constant at $k\lsim k_c\sim 1$ GeV \cite{MS,DKT,DMW,SS}. 
Phenomenologically the magnitude of $\alpha_{s,v}$ at $k\lsim k_c$ can be 
estimated from, say, the analysis of the heavy quark energy losses 
which gives \cite{DKT}
\beq
\int_{\mbox{0}}^{\mbox{\small 2 GeV}}\!dk\frac{\alpha_{s,v}(k)}{\pi}
\approx 0.36 \,\,
\mbox{GeV}\,.
\label{eq:2}
\eeq  
For the simplest prescription with frozen $\alpha_{s,v}$ at $k<k_c$ 
(the so-called $F$-model \cite{DKT}), using the one-loop expression
at $k>k_c$ one can obtain from (\ref{eq:2})
$\alpha_{s,v}(k<k_c)=\alpha_{s,v}^{fr}\approx 0.7$, and 
$k_c\approx 0.82$ GeV. These values are for $\Lambda_{QCD}=0.3$ GeV.

Unfortunately, at present there is no accurate 
information on $\alpha_s(k)$ for gluon emission from a 
fast parton in QGP.
Available pQCD calculations are performed in the static limit 
(see for example \cite{BPS,ESW,CH} and references therein). 
The running coupling constant obtained in \cite{ESW,CH} has a pole 
at $k/\Lambda_{QCD}\sim 3$ at $T\sim 250$ MeV. Thus, in pQCD, even 
for the static case, the situation with $k$-dependence of the 
in-medium running coupling constant at low $k$ is unclear. 
On the other hand, analysis of the lattice results within 
quasiparticle model gives the thermal $\alpha_s$ with a smooth 
$T$-dependence, and at $T\sim 250$ MeV $\alpha_s\approx 0.5$ \cite{LH}.
In the present paper in the absence of accurate information
on the in-medium running coupling constant for fast partons we 
perform calculations using the above $F$-model 
with different values of $\alpha_{s,p}^{fr}$.

{\bf 3. Evaluation of coherent medium correction to gluon spectrum.}
Let us now discuss technical aspects of our analysis of the CFSI.
We consider the gluon radiation from a fast quark 
(the generalization to the radiation from a gluon is trivial). 
We neglect multiple emission and consider only the leading order
splitting $q\rightarrow gq$. It is reasonable since the
effect is dominated by the gluons with small transverse momenta 
$k\lsim 1-2$ GeV. Note that we choose  the $z$-axis along the momentum of 
the initial fast parton, so for the central rapidity region in $AA$ 
collisions our $L$ is the ordinary transverse distance between the jet 
production point and the boundary of QGP.


We consider a fast quark with energy $E_i$ produced at $z=0$ which 
eventually splits 
at some $z>0$ into the gluon and final quark with the energies $E_g=xE_i$ 
and $E_f=(1-x)E_i$ respectively.
The corresponding matrix element can be written in the form
(below for simplicity we drop color factors)
\beq
T=i\int_{0}^{\infty}dz\int d\r g\bar{\psi}_{f}(\r,z)\gamma^{\mu}
A_{\mu}(\r,z)\psi_{i}(\r,z)\,,
\label{eq:3}
\eeq 
where $\psi_{i,f}(z,\r)$ are the wave functions of the initial and
final quarks, and $A_{\mu}$ is the wave function of the emitted gluon,
$\r$ is the transverse coordinate. In (\ref{eq:3}) we do not explicitly 
indicate 
the $z$ and $k$ dependence of the running coupling constant $g$.
We evaluate the matrix element (\ref{eq:3}) 
for small emission angles. Then, at high energies $E_j\gg m_q$ the quark 
wave functions using the ordinary light-cone spinor basis  
can be written as
\beq
\psi_j(\r,z)=\exp(iE_j z)\hat{U}_{j}\phi_j(z,\r)\,,
\label{eq:4}
\eeq
where the operator $\hat{U}_j$ reads 
\beq
\hat{U}_j=\left(\sqrt{2E_j}+\frac{\ab\pb+\beta m_q}{\sqrt{2E_j}}
\right)\chi_j\,.
\label{eq:5}
\eeq
Here $\chi_j$ is the quark spinor (normalized to unity), 
$\ab=\gamma^0\gb$, $\beta=\gamma^0$, and $\pb=-i\nabla_{\perp}$.
The gluon wave function can be written in a form 
similar to (\ref{eq:4}) (up to an obvious change of the spin operator). 
The transverse quark wave function $\phi_{j}(\r,z)$ entering (\ref{eq:4}) 
is governed
by the two-dimensional Schr\"odinger equation in which $z$ plays the role of
time
\beq
i\frac{\partial\phi_{j}(z,\r)}{\partial z}=
\frac{({\pb}^{2}+m^{2}_{q})}{2 E_{j}}\,\phi_{j}(z,\r)\,.
\label{eq:6}
\eeq
A similar equation holds for the gluon wave function.

Without a loss of generality we can take for the initial quark the 
plane wave state in the $\r$-plane and set $\pb_i=0$.
Then all the transverse wave functions can be written as
\beq
\phi_j(z,\r)=\exp\left\{i\left[\pb_j\r-\int_{0}^{z} d\xi
\frac{(\pb_{j}^{2}+m_{j}^{2}(\xi))}{2E_j}
\right]\right\}\,.
\label{eq:8}
\eeq
Eventually, the $\r$-integration in (\ref{eq:3}) will give 
$\pb_g+\pb_f=\pb_i=0$.
Note that since the quark mass is of only marginal significance 
(for light quarks) in the gluon radiation we will neglect the $z$-dependence of
$m_q$ and use the same quark mass in the vacuum and in QGP.
However, for the gluon transverse wave function
we use the $z$-dependent gluon mass: $m_g(z<L)=m_{g,p}$ and
$m_g(z>L)=m_{g,v}$.

Using the above formulas for the transition amplitude and wave functions 
with the help of the standard Fermi golden rule one can obtain after some
simple calculations for the gluon distribution (below $\kb=\pb_g$)
\beq
\frac{dN}{dxd\kb^{2}}=
\frac{dN^{(0)}}{dxd\kb^{2}}+
\frac{dN^{(1)}}{dxd\kb^{2}}\,,
\label{eq:9}
\eeq
\beq
\frac{dN^{(0)}}{dxd\kb^{2}}=\frac{C_F \alpha_s^v(k)}{\pi x}
\left(1-x+\frac{x^2}{2}\right)\frac{\kb^{2}}{(\kb^2+\mu_{v}^2)^2}\,\,,
\label{eq:10}
\eeq
\bea
\frac{dN^{(1)}}{dxd\kb^{2}}=\frac{2C_F \alpha_s^v(k)}{\pi x}
\left(1-x+\frac{x^2}{2}\right)
\left[1-\cos\left(\frac{(\kb^2+\mu_{p}^2)L}{2E_i x(1-x)}\right)\right]
\nonumber \\
\times
\frac{\kb^2 r(k) [(r(k)-1)\kb^2+r(k)\mu_{v}^2-\mu_{p}^2]}
{(\kb^2+\mu_{p}^2)^2 (\kb^2+\mu_{v}^2)}\,\,\,\,\,\,\,\,\,\,\,\,\,\,\,\,\,\,\,,
\label{eq:11}
\eea
where 
$\mu_i^2=m_q^2 x^2+m_{g,i}^2(1-x)\,
$,
$r(k)=
\sqrt{\alpha_{s,p}(k)/\alpha_{s,v}(k)}
$.
The first term on the r.h.s of (\ref{eq:9}) is the gluon 
spectrum in the vacuum, and the second one gives the medium correction of 
interest. In deriving (\ref{eq:10}), (\ref{eq:11}) we neglected
the small spin-flip contribution ($\propto m_q^2$).

Recall that the gluon formation length is
$
L_f\sim 2E_ix(1-x)/(\kb^{2}+\mu^2)
$
(we do not specify here the medium index since $\mu_{p}$ and $\mu_{v}$ 
are of the same order).
Thus we see that the argument of cosine in the r.h.s. of (\ref{eq:11}) is 
$\sim L/L_f$. Obviously,
at $L_f\ll L$ the rapidly oscillating cosine as function 
of $L$ will vanish upon averaging over the production point of the fast 
quark, and one gets $L$-independent correction to the vacuum term.
It can be written in the following physically transparent form 
\beq
\left.\frac{dN^{(1)}}{dxd\kb^{2}}\right|_{L_f\ll L}\approx
\left.\frac{dN^{(0)}}{dxd\kb^{2}}\right|_p
-
\left.\frac{dN^{(0)}}{dxd\kb^{2}}\right|_v
+
\frac{dN^{tran}}{dxd\kb^{2}}\,,
\label{eq:12}
\eeq
where the first two terms describe simply modification of the spectrum 
due to the change in $m_g$ and $\alpha_s$ (in an infinite QGP), and the last 
term is the contribution of the transition radiation which reads
\beq
\frac{dN^{tran}}{dxd\kb^{2}}=\frac{C_F \alpha_s^v(k)}{\pi x}
\left(1-x+\frac{x^2}{2}\right)
\left(
\frac{\kb r(k)}{\kb^2+\mu_{p}^2}-
\frac{\kb}{\kb^2+\mu_{v}^2}
\right)^2\,\,.
\label{eq:13}
\eeq
It can be derived from (\ref{eq:3}) taking for
the lower limit of the $z$-integral $-\infty$. Note that  
the change in $m_g$ and $\alpha_s$ both cause the transition radiation.

On the other hand, for the gluons with $L_f\gg L$ 
expanding cosine in (\ref{eq:11}) one gets the correction $\propto L^2$
\beq
\left.
\frac{dN^{(1)}}{dxd\kb^{2}}
\right|_{L_f\gg L}
\approx\frac{C_F \alpha_s(k)L^2}{\pi x}
\left(1-x+\frac{x^2}{2}\right)
\frac{\kb^2 r(k) [(r(k)-1)\kb^2+r(k)\mu_{v}^2-\mu_{p}^2]}
{4E_i^2 x^2(1-x)^2  (\kb^2+\mu_{v}^2)}\,\,.
\label{eq:14}
\eeq
In this regime one cannot separate the transition radiation.
Note that the above formulas demonstrate that the decrease of
the cutoff and coupling constant in QGP work in opposite directions.
One can also see that the relative contribution of the medium 
correction in (\ref{eq:9}) is larger for gluons with 
$L_f\lsim L$. These facts are consistent with intuitive expectation. 

{\bf 4. Numerical results.}
In numerical calculations we take: $m_{g,p}=0.4$ and $m_{g,v}=0.8$ GeV,
for the quark mass we take $m_{q,p}=m_{q,v}=0.3$ GeV.  
As was mentioned above in the absence of accurate information on 
the in-medium $\alpha_{s}$ at low $k$ we perform numerical calculations
parametrazing it in the same $F$-model as for $\alpha_{s,v}$ for
several values of $\alpha_{s,p}^{fr}$. To understand the sensitivity
of CFSI to $\alpha_{s,p}^{fr}$ we use for it
four values: 0.7, 0.5, 0.4, and 0.3. 
The first version corresponding to $\alpha_{s,p}=\alpha_{s,v}$ is 
unlikely to be realistic since it neglects the in-medium modification 
of $\alpha_s(k)$, but it allows to see the magnitude of the pure 
mass effect. The last value is also unlikely to be realistic. Indeed, 
the results of the analysis of the lattice data within the quasiparticle
model of QGP \cite{LH} say that $\alpha_s\approx 0.3$
occurs for $T\approx 2.2T_c$. Since the QGP can have a temperature 
above this value only during a short time in the initial stage of its 
evolution the value of $\alpha_{s,p}^{fr}=0.3$ is probably too
small for evaluation of the CFSI.
The value $\alpha_{s,p}^{fr}=0.5$ 
seems to be most reasonable. For instance, it is close  
the value $\alpha_s\approx 0.47$ obtained in \cite{LH} at 
$T\approx 1.5T_c\approx 250$ MeV.
Of course, it is not obvious that the thermal $\alpha_s$ can be 
extrapolated safely to higher energies. 
However, one can expect that in the splitting of fast partons the 
effect of thermostat on $\alpha_s$ can only be weaker than that for 
the thermal partons. If this is the case, the CFSI correction may be 
larger than our estimate. 

In Fig.~1 we show the $x$- dependence of the ratio 
$R(x)=\frac{dN}{dx}\left.\right/\frac{dN^{(0)}}{dx}$
for several $k$-windows at $E_i=40$ GeV averaged over $L$ in the
interval $[0,6]$ fm.
\begin{figure}[h]
\begin{center}
\input epsf
\epsffile{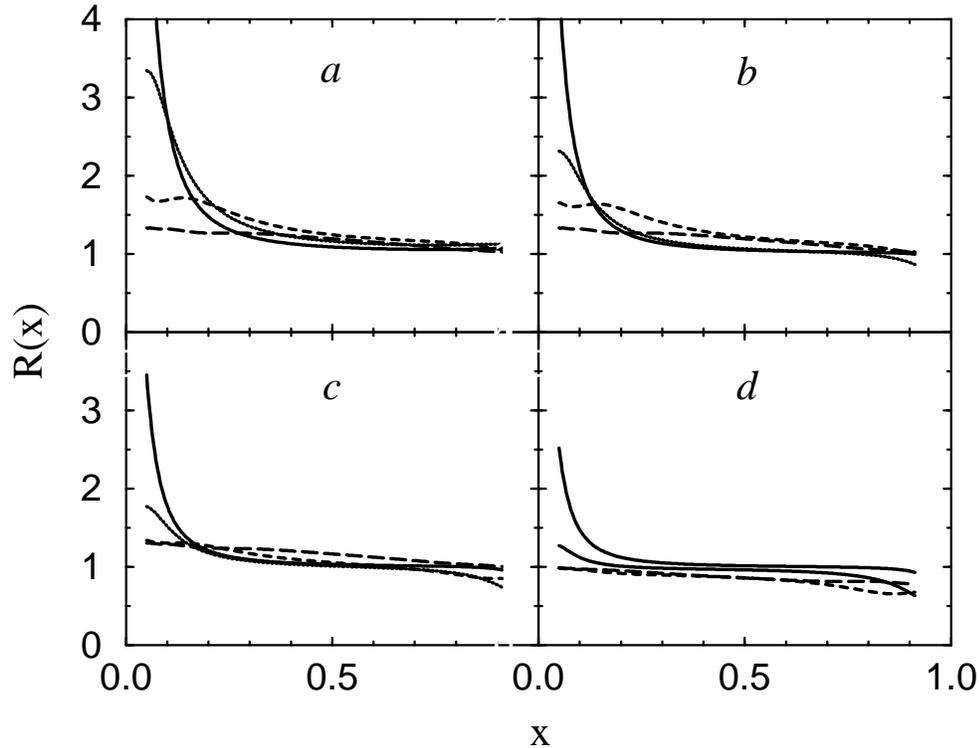}
\end{center}
\caption{
The ratio 
$R(x)=\frac{dN}{dx}\left.\right/\frac{dN^{(0)}}{dx}$ at $E_i=40$ GeV
for $\alpha_{s,p}^{fr}=$0.7 (a), 0.5 (b), 0.4 (c), and 0.3 (d)
evaluated using Eqs. (7)-(9) for the $k$-windows: [0,0.5] (solid line),
[0.5,1] (dotted line), [1,1.5] (dashed line), and [1.5,2] 
(long dashed line) GeV. The $x$-distributions obtained averaging over $L$ in
the interval [0,6] fm.
}      
\label{f1}
\end{figure}
As can be seen from Fig.~1 for $\alpha_{s,p}=\alpha_{s,v}$ the mass effect  
alone enhances considerably the probability of gluon emission at low $x$
and $k\lsim 1$ GeV. The curves for smaller values 
of $\alpha_{s,p}^{fr}$ show that the enhancement of the radiation due to
smaller cutoff in QGP is strongly compensated by 
the effect of smaller coupling constant in QGP, and for, probably unrealistic,
$\alpha_{s,p}^{fr}=0.3$ there is a kinematic region 
where CFSI suppresses the gluon radiation. 

Using (\ref{eq:11}) we also calculated the energy loss, defined as 
$$
\Delta E=E_i\int_{x_{min}}^{x_{max}} dx
\int_{0}^{k^{2}_{max}} d \kb^2
x\frac{dN^{(1)}}{dxd\kb^{2}} \,\,.
$$
For the limits of the $x$- and $\kb^2$-integration we take $x_{min}=m_g/E_i$,
$x_{max}=m_q/E_i$, and $k^2_{max}=\mbox{min}[E_i^2 x^2, E_i^2(1-x)^2]$.  
In Fig.~2 we show the results for $\Delta E$ as a function of $L$
at $E_i=$10, 20, 40, and 80 GeV.
\begin{figure}[h]
\begin{center}
\input epsf
\epsffile{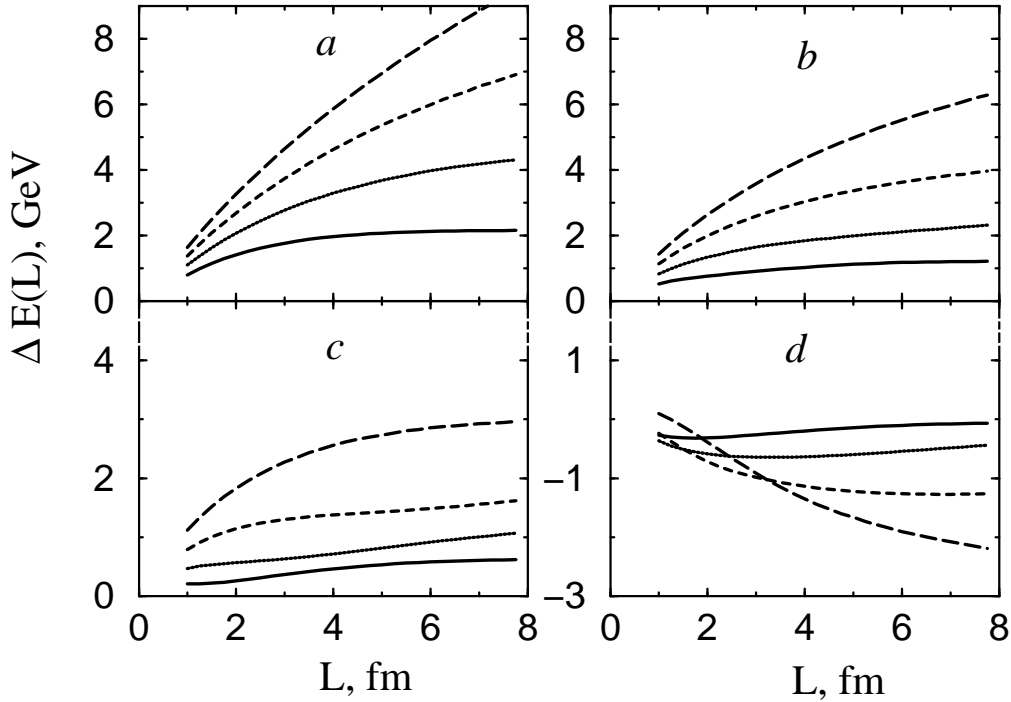}
\end{center}
\caption{
The quark energy loss $\Delta E$ due to the CFSI as a function of $L$ 
for $\alpha_{s,p}^{fr}=$0.7 (a), 0.5 (b), 0.4 (c), and 0.3 (d)
at $E_i=$10 (solid line), 20 (dotted line), 40 (dashed line), and 80 
(long dashed line) GeV.
}
\label{f2}
\end{figure}
As one can see for $\alpha_{s,p}^{fr}=$0.7, 0.5, and 0.4 the energy loss
is positive and rising with $L$ and $E_i$. 
For $\alpha_{s,p}^{fr}=0.3$ $\Delta E$ becomes negative which shows
that suppression of the gluon radiation due to a small coupling constant
becomes stronger than the enhancement caused by the mass effect.
This strong cancellation between the two competing effects
makes it difficult to give definitive predictions for the effect of CFSI
in the region $\alpha_{s,p}^{fr}\sim 0.3-0.4$.
Note that our $\Delta E$ for $\alpha_{s,p}^{fr}=0.5$ 
appears to be of the same order of magnitude as the GLV 
prediction \cite{GLV2} for the energy loss due to the induced radiation.

{\bf 5. Summary.}
We have shown that the final state interaction due to 
the change in the cutoff scale and running coupling
constant from the vacuum to QGP modifies the gluon radiation
from fast partons produced in $AA$-collisions.
The contribution to the energy loss of this mechanism may
be of the same order of magnitude as the induced gluon radiation.
However, accurate evaluation of the CFSI is a difficult task
since there are strong cancellations between the mass and running coupling
constant effects, and the results depend strongly
on the assumptions on the $k$-dependence of the in-medium 
$\alpha_{s}(k)$.

The results of the present paper raise a practical question, whether 
the jet tomographic analyses based on the theory of the induced 
gluon radiation can be used for extracting the density of hot QCD
medium produced in $AA$ collisions. At present, one cannot exclude
the possibility that the CFSI may appear to
be even more important than the induced radiation.
To clarify the situation it is highly desirable to study the influence of 
QGP on the running coupling constant for fast partons.

I am grateful to R. Baier and N.N. Nikolaev for discussions. I am also
grateful to J.~Speth for the hospitality
at FZJ, J\"ulich, where this work was completed.   
This work was partially supported by the grants INTAS
97-30494 and DFG 436RUS17/119/01.

\end{document}